\documentclass{article}

\font\sqi=cmssq8
\def\DR{\rm I\kern-1.45pt\rm R}
\def\DC{\kern2pt {\hbox{\sqi I}}\kern-4.2pt\rm C}
\newcommand{\ben}{\begin{enumerate}}
\newcommand{\een}{\end{enumerate}}
\newcommand{\beq}{\begin{equation}}
\newcommand{\eeq}{\end{equation}}
\newcommand{\bea}{\begin{eqnarray}}
\newcommand{\eea}{\end{eqnarray}}

\def\DH{\rm I\kern-1.5pt\rm H\kern-1.5pt\rm I}

\def\bbbz{{\mathchoice {\hbox{$\sf\textstyle Z\kern-0.4em Z$}}
{\hbox{$\sf\textstyle Z\kern-0.4em Z$}} {\hbox{$\sf\scriptstyle
Z\kern-0.3em Z$}} {\hbox{$\sf\scriptscriptstyle Z\kern-0.2em
Z$}}}}

\def\bbbz{{\rm I\!Z}} 
\chardef\ii="10
\begin{document}
\begin{center}
{\large\bf Hyperboloid, instanton, oscillator} \\
\vspace{0.4 cm} {\large Stefano Bellucci$^{1}$, Levon Mardoyan$^{2}$ and Armen
Nersessian$^{2}$ }\\
\vspace{0.2 cm} $^1${\it INFN-Laboratori Nazionali di Frascati,
Via E. Fermi 40,
00044 Frascati, Italy}\\
$^2${\it  Artsakh State University, Stepanakert} \& {\it Yerevan
State University, Yerevan, Armenia}\end{center}
\begin{abstract}
We suggest the exactly solvable model of the oscillator
 on a four-dimensional hyperboloid
 which  interacts with a
 $SU(2)$ instanton. We calculate its wavefunctions and spectrum.
\end{abstract}
\begin{center}
\end{center}
\setcounter{equation}0
\subsubsection*{Introduction}
There exists a well-known generalisation of the $N$-dimensional
isotropic oscillator  to spheres and hyperboloids, suggested by
Higgs \cite{higgs}. The uniqueness of this system is that it
inherits all constants of motion of the standard (flat)
oscillator, although its symmetry algebra becomes nonlinear
(wheras the symmetry algebra of the flat oscillator is $su(N)$).
However, this system does not respect the inclusion of external
gauge fields. For instance, the inclusion of a constant magnetic
field in the two-dimensional Higgs oscillator breaks even its
exact solvability. On  the other hand, a constant magnetic field
preserves the kinematical symmetries of a free particle on the
two-dimensional sphere and hyperboloid (which form, respectively,
$so(3)$ and $so(1.2)$ Lie algebras), as well as the exact
solvability of the planar (two-dimensional) oscillator. Similarly,
the inclusion of   the BPST instanton field preserves the
kinematical symmetries of a free particle moving on the four-
dimensional sphere and  the exact solvability of the
four-dimensional flat oscillator, but it breaks the exact
solvability of the Higgs oscillator on the four-dimensional
sphere. Instead, the analog of the oscillator on both complex
projective spaces $\DC P^N$ and their noncompact analogs, i.e.
Lobachevsky spaces ${\cal L}_N=SU(N.1)/U(N)$ \cite{cpn}, loosing
part of the hidden symmetries, remains, nevertheless, exactly
solvable in the presence of a constant magnetic field. Since $\DC
P^1$ is the two-dimensional sphere, and ${\cal L}_1$ is the
two-dimensional hyperboloid, the above model seems to be an
appropriate alternative to the two-dimensional Higgs oscillator,
as the analog of the planar oscillator with constant magnetic
field. These systems are defined by the potential \cite{cpn}
 \beq V_{\DC
P^N}=\omega^2r^2_0\frac{u_a\bar u_a}{u_0\bar u_0},\qquad u_0\bar
u_0+\epsilon u_a\bar u_a=r^2_0,\quad a=1,\ldots N,\quad
\epsilon=\pm 1 \label{bn}\eeq
 where $u_a/u_0$ are
homogeneous complex coordinates for  $\DC P^N$ and ${\cal L}_N$,
$\epsilon=1 $ corresponds to the ${\DC P}^N$,
 and   $\epsilon=-1 $ corresponds to the ${\cal L}_N$.

In \cite{lectures} it was claimed, that the potential (\ref{bn}),
would be the appropriate generalization of the oscillator on the
quaternionic projective spaces with $\DH P^N$ respecting the
inclusion of the (constant uniform) instanton field, provided we
interpret ${\bf u}_a$ and ${\bf u}_0$ as quatenionic coordinates
of the ambient quaternionic space $\DH^{N+1}$. For $\DH P^1$ (i.e.
for the four-dimensional sphere) it was shown that this is indeed
the case \cite{su2}. Moreover,  in contrast with the case of $\DC
P^1$, the spectrum of the system depends on the topological charge
of the instanton (what might be connected to the behaviour of
two-dimensional noncommutative quantum mechanics models in a
constant magnetic field \cite{nc}). The invention and study of
this model was motivated by the recently suggested theory of the
four-dimensional Hall effect \cite{4hall}, which attracted a wide
interest in the physics community (see, e.g.
\cite{4hallothers,polchinski}). This theory is based on the
quantum mechanics of colored particle moving on a four-dimensional
sphere in the field of a $SU(2)$ Yang monopole \cite{yang} (which
is equivalent to the BPST instanton \cite{bpst}). The inclusion of
the potential reduced it to the effective three-dimensional edge
theory. The key role in this model is played by the external
instanton field, which provides it with a degenerate ground state,
becoming infinite in the planar limit. This theory displays a few
interesting phenomena, such as an infinite gapless sequence of
massless particle excitations with any spin. The initial (without
potential term) symmetry of this theory is $SO(5)$, and it has no
relativistic interpretation.
 On the other hand,  the quantum mechanics
 of the colored particle on the four-dimensional hyperboloid,
 moving in the (constant uniform) field of $SU(2)$ instanton
  would possess a $SO(4.1)$ symmetry. There is no doubt that it will have a
  degenerate ground state, hence, developing the theory of
  four-dimensional quantum Hall effect will be possible on this
 quantum-mechanical background too.
One can expect that this hypothetic, $SO(4.1)$-symmetric theory
 of the four-dimensional Hall effect will have a
relativistic interpretation. By this reason, the construction of
the non-compact analog of the model \cite{su2}, i.e. the
exactly-solvable model of the oscillator on the four-dimensional
hyperboloid interacting with the SU(2) instanton field, seems to be
even more important, than the initial ``compact" system. This
is the subject of the present paper.

The paper is arranged as follows. In  {\sl  Section 1}  we
construct the hyperbolic analog of the BPST instanton and suggest the
appropriate oscillator potential. In {\sl Section 2} we get
the energy spectrum and wavefunctions of the system.

\subsubsection*{1.Instanton}

In this Section we construct the SU(2) instanton and
anti-instanton on the four-dimensional two-sheet hyperboloid, which
defines the field configuration with constant magnitude, i.e.
the hyperbolic analog of the BPST instanton. Also, we present the
hyperbolic analog of the oscillator potential on the
four-dimensional sphere considered in Ref.\cite{su2}.
 It is convenient to get these basic
ingredients of the model by the use of quaternions, following
\cite{atiah}.

Let us parameterize the (ambient ) pseudo-Euclidean space
$\DR^{4.1}$ by the real coordinate $x_0$ and the quaternionic one
${\bf x}=x_4+\sum_{a=1}^3 x_a{\bf e}_a$,
 with ${\bf e}_a{\bf e}_b=-\delta_{ab}+\varepsilon_{abc}{\bf e}_c $,
  $\bar{\bf e}_a= -{\bf e}_a$. Notice that $t_a\equiv {\bf
  e}_a/2$ form a $su(2)$ algebra: $[t_a,t_b]=\varepsilon_{abc}t_c $.
 In
terms of these coordinates the metric on $\DR^{4.1}$ reads \beq ds^2=d{\bf x} d{\bf
\bar x}-dx^2_0. \label{methric0}\eeq
Imposing the constraint \beq
x^2_0-{\bf x}{\bf \bar x}=r^2_0, \eeq
we shall get the metric on the
four-dimensional hyperboloid.
 It is convenient to resolve this constraint, choosing
 \beq x_0=r_0\frac{1+{\bf
w}{\bf\bar w}}{1-{\bf w}{\bf\bar w}}, \quad {\bf x}=r_0\frac{2{\bf
w}}{1-{\bf w}{\bf\bar w}}, \eeq where $|{\bf w}|<1$ for the upper
sheet of hyperboloid, and $|{\bf w}|>1$ for the lower one. In
these coordinates the metric (\ref{methric0}), restricted to the
hyperboloid, looks as follows:
 \beq
 ds^2=\frac{4r^2_0 d{\bf w}d{\bf\bar w}}{(1-{\bf w}{\bf\bar w})^2}.
 \label{metric2}\eeq
This is precisely the quaternionic analog of the Poincar\'e model
of the Lobachevsky plane (two-dimensional two-sheet hyperboloid).
The instanton and anti-instanton solutions are defined by the
following expressions:
 \beq {\bf A}_{+}={\rm Im}\frac{{\bf w}d{\bf\bar
w}}{{\bf w\bar w}-1}=-{\rm Im}\frac{{\bf x}d{\bf\bar
x}}{2r_0(x_0+r_0)}, \quad {\bf A}_{-}= {\rm Im}\frac{{\bf \bar w}
d {\bf w}}{{\bf w\bar w}-1}=-{\rm Im}\frac{{\bf \bar x}d{\bf
x}}{2r_0(x_0+r_0)}.\label{ac} \eeq Let us prove it following the
arguments of Atiah \cite{atiah}. In quaternionic notation the
strength of the $SU(2)$ gauge field with potential ${\bf A}={A_a}{\bf
e}_a/2$ is defined by the expression ${\bf F}=d{\bf A}+{\bf
A}\wedge {\bf A}$. Hence, calculating it for the ${\bf A}_\pm$ ,
we get \beq {\bf F}_+=-\frac{d{\bf w}\wedge d{\bf \bar w}}{({\bf
w\bar w}-1)^2}=\frac{{\bf e}_a\widetilde
{\omega}^{(2)+}_a}{2(1-{\bf w\bar w})^2},\quad {\bf
F}_-=-\frac{d{\bf \bar w}\wedge d{\bf w}}{({\bf w\bar
w}-1)^2}=\frac{{\bf e}_a\widetilde {\omega}^{(2)-}_a}{2(1-{\bf
w\bar w})^2}, \label{Fpm}\eeq where
 \beq {{\widetilde\omega}^{(2)\pm}_a}=(\pm dw_4\wedge dw_1 +
 dw_2\wedge dw_3,\;\pm dw_4\wedge dw_2 +
 dw_3\wedge dw_1,\;\pm dw_4\wedge dw_3 +
 dw_1\wedge dw_2,\;)
 \eeq
The set  $\widetilde\omega^+_a$ defines, precisely,
 the basis of  self-dual two-forms, and $\widetilde\omega^-_a$ that of
anti-self-dual ones.
Consequently, the ${\bf
F}_+$  is a self-dual field, and  ${\bf
F}_-$is a anti-self-dual one: $^\star {\bf
F}_\pm =\pm {\bf F}_\pm $.

 Seemingly, $|{\bf F}^\pm | \to \infty$, when $|w|\to \infty$ (and
$|{\bf x}| \to \infty $). However, considering the classical motion of
a particle the on four-dimensional hyperboloid in the presence of
these  fields (see \cite{duval}), one can see that the
constructed instanton and anti-instanton configurations have a
constant uniform magnitude. Indeed, the magnitude of the gauge
field is defined as the strength multiplied on the inverse
metrics. Hence, the product of the matrices
 ${\bf F}_\pm $ given by (\ref{Fpm})
 on the inverse to the metrics  (\ref{metric2}) is equal to
 the constant two-form
 $-{\bf e}_a{\widetilde\omega}^\pm_a/8r^2_0$.

Finally, let us conclude this section writing down the components
of connections (\ref{ac}) in real coordinates
%
 \bea & {A}^{\pm}_1&=2\frac{\pm w_4dw_1
+w_3dw_2-w_2dw_3\mp w_1dw_4}{w_\mu w_\mu-1}=-\frac{\pm x_4dx_1
+x_3dx_2-x_2dx_3\mp x_1dx_4}{r_0(x_0+r_0)}\\[4mm]
&{A}^{\pm}_2&=2\frac{- w_3dw_1 \mp w_4dw_2-w_1dw_3\pm
w_2dw_4}{w_\mu w_\mu-1}=-\frac{ -x_3dx_1
\mp x_4dx_2-x_1dx_3\pm x_2dx_4}{r_0(x_0+r_0)}\\[4mm]
& {A}^{\pm}_3&=2\frac{ w_2dw_1 - w_1dw_2 \mp w_4dw_3 \pm
w_3dw_4}{w_\mu w_\mu-1}= -\frac{ x_2dx_1 - x_1dx_2\mp x_4dx_3\pm
x_3dx_4}{r_0(x_0+r_0)}
 \eea

\subsubsection*{2.Oscillator}
The quantum  mechanics  of the colored particle moving on the
four-dimensional hyperboloid in the presence of the potential $V$ and
(anti)instanton field is described  by the Hamiltonian \beq
{\widehat H}= -\frac{\hbar^2(1-w_\mu w_\mu )^4}{2r^2_0 }{\cal
D}_\mu (1-w_\mu w_\mu )^{-2}{\cal D}_\mu  +V(w,\bar w),\quad {\cal
D}_\mu =\partial/\partial{w_\mu}+iA_{(a)\mu}T_a
 \eeq
where  $T_a$ are the $SU(2)$ generators on the internal space
$S^2$ of the (anti)instanton, $[\hat T_a,\hat T_b] =
i\epsilon_{abc}\hat T_c $, and $A^a_A$ is defined by (\ref{ac}).

We choose the
 oscillator potential on the four-dimensional hyperboloid given by the expression
 \beq
 V_{osc}= {2\omega^2r^2_0} {\bf w\bar w}=
 {2\omega^2r^2_0}\frac{x_0-r_0}{x_0+r_0},
 \label{pot+}\eeq
which is similar to the one on the  complex projective space $\DC P^1$ \cite{cpn},
 and on the quaternionic projective spaces  $\DH P^1$ \cite{lectures,su2}.
Let us show that this oscillator system is an exactly solvable one
  and calculate its wavefunction and spectrum.

We restrict ourselves to the upper sheet of hyperboloid
 and introduce the ``hyperspherical" coordinates \beq x_0 = r_0\cosh
\theta ,\; x_2 + ix_1 = r_0 \sinh \theta \sin \frac{\beta}{2}{\rm
e}^{i\frac{\alpha -\gamma}{2}}, \; x_4 + ix_3 = r_0 \sinh \theta
\cos \frac{\beta}{2}{\rm e}^{i\frac{\alpha + \gamma}{2}},\eeq or,
equivalently, \beq w_2+iw_1=\tanh\frac{\theta}{2}\sin
\frac{\beta}{2}{\rm e}^{i\frac{\alpha -\gamma}{2}} \quad
w_4+w_3=\tanh\frac{\theta}{2}\cos \frac{\beta}{2}{\rm
e}^{i\frac{\alpha + \gamma}{2}}  \eeq
 where $\theta\in [0,\infty )$,
$\beta \in [0,\pi]$, $\alpha \in [0,2\pi)$, $\gamma \in [0,4\pi)$.

In these terms the quantum Hamiltonian with the oscillator
potential  (\ref{pot+}) and with the instanton field $A^+_a$
 reads
 \beq {\cal
H}^+=-\frac{1}{2r^2_0}\left[ \frac{1}{ \sinh^3
\theta}\frac{\partial}{\partial \theta} \left(\sinh^3 \theta
\frac{\partial}{\partial \theta}\right) +\frac{2{\hat
L}^2}{1-\cosh\theta} + \frac{2{\hat J}^2}{1+\cosh\theta}\right]+
{2\omega^2r^2_0}\frac{\cosh\theta-1}{\cosh\theta+1}.
\label{qH+}\eeq Here ${\hat L}_a$ are the components of the $SU(2)$
momentum $[\hat L_a,\hat L_b] = i\epsilon_{abc}\hat L_c $,
\beq
{\hat L}_1 = i\left(\cos{\alpha}\cot{\beta}
\frac{\partial}{\partial \alpha}+
\sin{\alpha}\frac{\partial}{\partial \beta} -
\frac{\cos{\alpha}}{\sin{\beta}}
\frac{\partial}{\partial \gamma}\right), \;
{\hat L}_2 = i\left(\sin{\alpha}\cot{\beta}
\frac{\partial}{\partial \alpha} -
\cos{\alpha}\frac{\partial}{\partial \beta} -
\frac{\sin{\alpha}}{\sin{\beta}}
\frac{\partial}{\partial \gamma}\right) ,\;
{\hat L}_3 = -i\frac{\partial}{\partial \alpha}.
\label{L}\eeq
and
$\hat J_a = \hat L_a + \hat T_a$,
\beq
[\hat L_a,\hat L_b] = i\epsilon_{abc}\hat L_c,\,\,\,\,\,\,
[\hat L_a,\hat J_b] = i\epsilon_{abc}\hat L_c,\,\,\,\,\,\,
[\hat J_a,\hat J_b] = i\epsilon_{abc}\hat J_c,
\eeq
It is convenient to represent the generators $T^a$
in terms of $S^3$, as in  (\ref{L}) (where, instead of $\alpha,\beta,\gamma$,
there appear the coordinates of $S^3$, $\alpha_T,\beta_T,\gamma_T$), with the
following condition imposed
\beq {\hat
T}^2\Psi(\alpha,\beta,\gamma,\theta,\alpha_T,\beta_T,\gamma_T)=
T(T+1)\Psi
(\alpha,\beta,\gamma,\theta,\alpha_T,\beta_T,\gamma_T),\eeq which
corresponds to the fixation of the isospin $T$.
 Notice that the
generators $\hat J_a$, $\hat L^2$, $\hat T^2$ are constants of
motion, while $\hat L_a$, $\hat T_a$ do not commute with the
Hamiltonian.

In order to  solve the  Schr\"odinger equation
${\cal H}\Psi={\cal E}\Psi$, we introduce the separation Ansatz
\beq
\Psi(\theta,\alpha,\beta,\gamma, \alpha_T,\beta_T,\gamma_T) =
Z(\theta )\Phi(\alpha,\beta,\gamma,\alpha_T,\alpha_T, \gamma_T).
\label{sep}\eeq
where $\Phi$ are the eigenfunctions of ${\hat L}^2$, ${\hat T}^2$ and
${\hat J}^2$ with the eigenvalues $L(L+1)$, $T(T+1)$ and $J(J+1)$.
Thus, $\Phi$ can be represented in the form
\beq
\Phi =
\sum_{M=m+t}\left(JM|L,m';T,t'\right)D_{mm'}^L(\alpha,\beta,\gamma)
D_{tt'}^T(\alpha_T,\beta_T,\gamma_T)
\eeq
where $\left(JM|L,m';T,t'\right)$ are the Clebsh-Gordan coefficients
and $D_{mm'}^L$ and $D_{tt'}^T$ are the Wigner functions.

Using  the above separation Ansatz, we get
 the following``radial" Schr\"odinger equation:
\begin{eqnarray}
\frac{1}{\sinh^3\theta}\frac{d}{d\theta}\left(\sinh^3\theta
\frac{dZ}{d\theta}\right)+\frac{2J(J+1)}{1+\cosh\theta }Z+
\frac{2L(L+1)}{1-\cosh\theta }Z+ 2r^2_0\left( {\cal
E}-2\omega^2r^2_0\frac{\cosh\theta -1}{\cosh\theta +1}\right)Z=0,
\label{2.4.1}
\end{eqnarray}
Now, making the substitution $Z(\theta )=\sinh^{-3/2}\theta
R(\theta )$, we end up with the equation \beq \frac{d^2 R}{d\theta^2} +
\left[\varepsilon -\frac{{ L}({ L} +1)+3/16}{\sinh^2\theta/2}
+\frac{{\widetilde J}({\widetilde J}
+1)+3/16}{\cosh^2\theta/2}\right]R=0, \eeq where we introduced the
notation \beq \varepsilon=2r^2_0{\cal
E}-4\omega^2r^4_0-\frac{9}{4},\qquad {\widetilde J}({\widetilde
J}+1)\equiv J(J+1)+4\omega^2r^2_0. \label{epsilon}\eeq The same
equation appears also in the Schroedinger equation of the Higgs
oscillator on a $4-$dimensional hyperboloid \cite{georgeN}.

The regular solution of this equation is the hypergeometric function
\beq R_{nJL}=(\sinh\frac{\theta}{2})^{2L+3/2}
(\cosh\frac{\theta}{2})^{2n-2{\widetilde J}-1/2}{_2}F_1(-n,n
+2{\widetilde J }+1,2L+2,\tanh^2\theta/2), \eeq
 where
 $n= \sqrt{-\varepsilon }+{\widetilde J}-L-1/2$
  is a non-negative integer number
 $n=0, 1,2,\ldots, [{\widetilde J}-L-1/2]$.
Taking into account the expression (\ref{epsilon}), we get the
energy spectrum of the system \beq {\cal E}= \frac{({\widetilde J
}-J)(n+L+1)}{r^2_0} -\frac{(n+L-J-1)(n+ L-J+2)}{2r^2_0}.
 \label{ener}\eeq
The  regular normalized wavefunction is defined by the expression \beq
Z(\theta)= \sqrt{\frac {(2{\widetilde
J}-2L-2n-1)(n+2L+1)!\Gamma(2{\widetilde J}-n+1)}
{n!\Gamma(2{\widetilde J} -2{ L}-n)}}\left(\sinh
\theta\right)^{-3/2} R_{nJL} \;. \label{wf}\eeq
 Let us remind
that \beq J=|L-T|,|L-T|+1,...,L+T,\qquad
n=0, 1,2,\ldots, [{\widetilde J}-L-1/2],
 \qquad L=0, {1}/{2}, 1,\ldots
\label{LJdef}\eeq
In the absence of the instanton field  one has
${\widetilde J}=J$.
 In this
case one can introduce the principal quantum number ${\cal
N}=n+J+L$, and get  standard expressions for the  spectrum and
  wavefunctions of the oscillator on the
four-dimensional hyperboloid.\\

{\bf Remark 1.} A similar system with the anti-instanton  field is
described by the Hamiltonian
 \beq {\cal
H}^-=\frac{1}{2r^2_0}\left[ \frac{1}{ \sinh^3
\theta}\frac{\partial}{\partial \theta} \left(\sinh^3 \theta
\frac{\partial}{\partial \theta}\right) +\frac{2{\hat
L}^2}{1+\cosh\theta} + \frac{2{\hat J}^2}{1-\cosh\theta}\right]+
{2\omega^2r^2_0}\frac{\cosh\theta-1}{\cosh\theta+1}.
\label{qH-}\eeq Hence, its spectrum  and wavefunctions
 can be
obtained from the above ones, (\ref{ener}),(\ref{wf}) by the
redefinition \beq
{\widetilde L}\to {\widetilde J} , \quad J\to L,
\qquad {\widetilde J}({\widetilde J}+1)\equiv
J(J+1)+2\omega^2r^2_0
  \; . \label{ma0}\eeq

{\bf Remark 2.} The above results could be easily extended to the
system with the ``singular oscillator" potential defined as
follows: \beq V^{so}= 2\omega^2r^2_0\frac{\cosh\theta -1
}{\cosh\theta + 1}+ 2\omega^2_1r^2_0\frac{\cosh\theta + 1
}{\cosh\theta -1}. \label{pot-}\eeq

The spectrum and wavefunctions of this system with the instanton field
 can be obtained from (\ref{ener}), (\ref{wf}) by the
redefinition \beq
 J\to {\widetilde J},
\qquad {\widetilde J}({\widetilde J}+1)\equiv
J(J+1)+2\omega^2_1r^2_0,\quad {\cal E}\to {\cal
E}-2\omega^2_1r^2_0
  \; . \label{ma0}\eeq
  Similarly, for the anti-instanton configuration we should
  make the following substitution:
\beq
 L\to {\widetilde L},
\qquad {\widetilde L}({\widetilde L}+1)\equiv
L(L+1)+2\omega^2_1r^2_0,\quad {\cal E}\to {\cal
E}-2\omega^2_1r^2_0
  \; . \label{ma1}\eeq
Hence, the instantonic singular oscillator with ``characteristic
frequencies" $(\omega, \omega_1 )$ is ``isomorphic"  to the
anti-instantonic singular oscillator with ``characteristic
frequencies" $(\omega_1, \omega )$\\

%
%
\subsubsection*{3.Discussion}
We constructed the hyperbolic analogs of the BPST (anti)instanton
and of the oscillator potential,  which preserve the exact
solvability of the particle moving on a four-dimensional
hyperboloid in their presence. We calculated the energy spectrum
and the wavefunctions of this  model and found, that it possesses
a degenerate ground state. Hence, we suggest that this system
could form the appropriate ground for developing the relativistic
theory of the higher-dimensional Hall effect. The system inherits
the asymmetry with respect to instanton and anti-instanton fields,
earlier observed in the models on the four-dimensional plane
\cite{polchinski} and sphere \cite{su2}. Also, it has a finite
discrete energy spectrum, which is typical for the systems on
spaces with constant negative curvature. Notice that the suggested
 system  could be viewed as a spherical part of the
 quantum-mechanical system on the $\DR^{4.1}$ describing the motion of a
 particle interacting with the
 ``hyperbolic Yang monopole" (\ref{ac}) and potential (\ref{pot+}),
 where $r_0$ is a dynamical variable.
 Hence, it could by obtained, by the reduction associated with the second Hopf map,
from the appropriate  systems on $\DR^{4.4}$ and on the ${\cal
L}_3=SU(3.1)/U(2)$ (compare, respectively, with  \cite{mardoyan})
and \cite{casteill}), which
  are specified by the absence of external gauge fields.
  Finally, let us mention that there is a kind of
  duality between monopoles and relativistic spinning particles
  (at the moment it is part of a folklore in physics, but probably for the
  first time it was pointed out in \cite{balachadran}). From this viewpoint
    the non-relativistic particle  moving on the
    hyperboloid in the presence of an instanton field is dual
to the (4+1)- dimensional massive spinning particle, similarly to
the duality between a non-relativistic  particle moving on the
two-dimensional hyperboloid in the presence of a constant magnetic
field, and the free massive relativistic (2+1)- dimensional
particle \cite{anyon}).
A possible direction for future developments is the consideration of supersymmetric
extensions, as it was done in the
first reference \cite{cpn} and \cite{susy}.\\
\\

{\large Acnowledgments.} We would like to thank Francisco Morales,
Ruben Mkrtchyan and Ruben Poghossian for valuable discussions and
useful remarks. The work of S.B. has been supported in part by the European Community Human
Potential Program under contract MRTN-CT-2004-005104 ``Constituents,
fundamental forces and symmetries of the universe''.
L.M. and A.N. are partially supported by the
NFSAT-CRDF grant ARPI-3228-YE-04.


\begin{thebibliography}{99}
\bibitem{higgs}P.~W.~Higgs,
 J.\ Phys.{\bf A12} (1979) 309;

H.~I.~Leemon,
J.\ Phys. {\bf A12} (1979) 489.
\bibitem{cpn}S.~Bellucci, A.~Nersessian,
 Phys.\ Rev. {\bf D67} (2003) 065013,
[Erratum-ibid. {\bf D 71}(2005) 089901];

  S.~Bellucci, A.~Nersessian, A.~Yeranyan,
  Phys.\ Rev.\ D {\bf 70} (2004) 085013;

S.~Bellucci, A.~Nersessian, A.~Yeranyan,
Phys.\ Rev.\ D {\bf 70} (2004) 045006.

\bibitem{lectures}A. Nersessian, ``Elements of (super-)Hamiltonian formalism",
Lecture Notes in Physics [arXiv:hep-th/0506170].
\bibitem{su2}L.~Mardoyan, A.~Nersessian, 
 Phys.\ Rev.\ {\bf B72} (2005) 233303.  


\bibitem{nc}
 S.~Bellucci, A.~Nersessian, C.~Sochichiu,
  Phys.\ Lett.\ B {\bf 522} (2001) 345
  [arXiv:hep-th/0106138];

S.~Bellucci, A.~Nersessian,
Phys.\ Lett. {\bf B542} (2002) 295
[arXiv:hep-th/0205024];
%

S.~Bellucci,
Phys. Rev. {\bf D67} (2003) 105014
[arXiv:hep-th/0301227].

\bibitem{4hall}S.~C.~Zhang, J.~P.~Hu,
Science {\bf 294} (2001) 823.
%
%
%
  \bibitem{4hallothers}
B.~A.~Bernevig, C.~H.~Chern, J.~P.~Hu, N.~Toumbas, S.~C.~Zhang,
 Annals Phys.\  {\bf 300} (2002) 185;

M.~Fabinger, 
JHEP {\bf 0205} (2002) 037;


D.~Karabali,  V.~P.~Nair, 
 Nucl.\ Phys.{\bf B641}(2002) 533;

V.~P.~Nair,  S.~Randjbar Daemi, 
Nucl.~Phys.~{\bf B679} (2004) 447 [arXiv:hep-th/0309212];

A.~P.~Polychronakos,
 Nucl.Phys.{\bf B705} (2005) 457.
%
%

\bibitem{polchinski}
H.~Elvang,  J.~Polchinski, ``The quantum Hall effect on $R^4$'',
[arXiv:hep-th/0209104].

\bibitem{atiah}M.~F.~Atiah, {\sl Geometry of the Yang-Mills Fields},
Accademia Nazionale dei Lincei, Scuola Normale Superiore, Lezioni
Ferminale, Pisa, 1979.

\bibitem{yang}  C.~N.~Yang,
  J.\ Math.\ Phys.\  {\bf 19} (1978) 320.
  J.\ Math.\ Phys.\  {\bf 19} (1978) 2622.
\bibitem{bpst}A.~A.~Belavin, A.~M.~ Polyakov, A.~S.~Schwarz,
Yu.~S.~Tyupkin, Phys.\ Lett. {\bf B59} (1975) 85.
\bibitem{duval}C.~Duval and P.~Horvathy,
  Annals Phys.\  {\bf 142} (1982) 10.

\bibitem{georgeN}E.~G.~Kalnins, W.~J.~Miller and G.~S.~Pogosyan,
  Phys.\ Atom.\ Nucl.\  {\bf 65} (2002) 1086.


\bibitem{casteill}
  S.~Bellucci, P.~Y.~Casteill,  A.~Nersessian,
  Phys.\ Lett.\ B {\bf 574} (2003) 121.
  [arXiv:hep-th/0306277].
\bibitem{mardoyan}
L.~G.~Mardoyan, A.~N.~Sissakian, V.~M.~Ter-Antonyan,
 Phys.\ Atom.\ Nucl.\ {\bf 61} (1998) 1746.
\bibitem{balachadran} M.~V.~Atre, A.~P.~Balachandran and T.~R.~Govindarajan,
  Int.\ J.\ Mod.\ Phys.\ A {\bf 2} (1987) 453.



\bibitem{anyon}M.~S.~Plyushchay,
  Mod.\ Phys.\ Lett.\ A {\bf 10} (1995) 1463;

A.~Nersessian, Mod.\ Phys.\ Lett.\  {\bf A12} (1997), 1783.

\bibitem{susy} S. Bellucci, A. Nersessian, "Supersymmetric Kaehler oscillator in a constant magnetic field,"
[arXiv:hep-th/0401232].

\end{thebibliography}
\end{document}